\documentclass[prd, amsfonts, twocolumn, nofootinbib, showpacs]{revtex4}
\usepackage{graphicx}
\usepackage{epsfig}
\usepackage{color}
\usepackage{amsmath}
\usepackage{amssymb}
\newcommand{\be}{\begin{equation}}
\newcommand{\ee}{\end{equation}}
\newcommand{\bea}{\begin{eqnarray}}
\newcommand{\eea}{\end{eqnarray}}

\newcommand{\gapp}{\mathrel{\raise.3ex\hbox{$>$}\mkern-14mu
\lower0.6ex\hbox{$\sim$}}}
\newcommand{\lapp}{\mathrel{\raise.3ex\hbox{$<$}\mkern-14mu
\lower0.6ex\hbox{$\sim$}}}
\def\bbox{{\,\lower0.9pt\vbox{\hrule \hbox{\vrule height 0.2 cm
\hskip 0.2 cm \vrule  height 0.2 cm}\hrule}\,}}

\begin{document}
\title{Observing a wormhole }
\author{De-Chang Dai\footnote{corresponding author: De-Chang Dai$^{1,2}$,\\ email: diedachung@gmail.com\label{fnlabel}}, Dejan Stojkovic$^3$,}
\affiliation{$^1$ Center for Gravity and Cosmology, School of Physics Science and Technology, Yangzhou University, 180 Siwangting Road, Yangzhou City, Jiangsu Province, P.R. China 225002 }
\affiliation{ $^2$ Department of Physics, Case Western Reserve University,
10900 Euclid Avenue, Cleveland, OH 44106 }
\affiliation{ $^3$ HEPCOS, Department of Physics, SUNY at Buffalo, Buffalo, NY 14260-1500, U.S.A.}


\begin{abstract}
\widetext
If a traversable wormhole smoothly connects two different spacetimes, then the flux cannot be separately conserved in any of these spaces individually. Then objects propagating in a vicinity of a wormhole in one space must feel influence of objects propagating in the other space. We show this in the cases of the scalar, electromagnetic, and gravitational field.
The case of gravity is perhaps the most interesting. Namely, by studying the orbits of stars around the black hole at the center of our galaxy, we could soon tell if this black hole harbors a traversable wormhole. In particular, with a near future acceleration precision of $10^{-6} m/s^2$, a few
solar masses star orbiting around Sgr A* on the other side of the wormhole at the distance of a few gravitational radii would leave detectable imprint
on the orbit of the S2 star on our side. Alternatively, one can expect the same effect in black hole binary systems, or a black hole - star binary systems.
Another result that we find very interesting is that gravitational perturbations can be felt even on the other side of the non-traversable wormhole.

\end{abstract}

\maketitle

\section{ Introduction}
Wormholes have been always attracting a lot of attention for various reasons ranging from pure academic interests and science fiction to a possible explanation of entanglement of particles in quantum mechanics \cite{ER,geons,wheeler,Baum:1984mc,Hawking:1984hk,Coleman:1988tj,gibbons,Morris:1988cz, Morris:1988tu,visser,Maldacena:2013xja,Gratus:2019ned,Bronnikov:2019gsr,Bronnikov:2018vbs,Ovgun:2018xys,Jusufi:2017mav,Jusufi:2016leh,Krasnikov:2008kr}.  The purpose of this work, however, is to establish a clear link between wormholes and astrophysical observations.
By definition, a wormhole smoothly connects two different spacetimes. If the wormhole is traversable, then the flux (scalar, electromagnetic, or gravitational) can be conserved only in the totality of these two spaces, not individually in each separate space. Suppose that there is a physical electric charge on one side of the wormhole. An observer on the other side where there is no physical electric charge sees the electric flux coming out of the wormhole, so he concludes that the wormhole is a charged object. Any measurement that he can perform by measuring the flux would tell him that the wormhole contains charge (though there is no real charge at the wormhole). In other words the flux is only apparently conserved in each space separately, but strictly conserved only if we consider the entirety of both spaces.
 A time dependent gravitational case is even more indicative.  If a real gravitational source is time dependent (e.g. a star orbiting a wormhole mouth), an observer on the side where there is no source will conclude that the gravitational perturbations he is observing cannot be sourced by a static wormhole.

As a direct consequence, trajectories of objects propagating in a vicinity of a wormhole must be affected by the distribution of masses/charges in the space on the other side of the wormhole. Since wormholes in nature are expected to exist only in extreme conditions, e.g. around black holes, the most promising systems to look for them are either large black holes in the centers of galaxies, or binary black hole systems.  We study motion of a star S2 which orbits a super massive black hole in Sgr A*  at the center of our galaxy and demonstrate that the near future data will be able to tell us if this black hole harbors a wormhole.

\section{  Charged particle and a wormhole in a flat space}
\begin{figure}
\includegraphics[width=6cm]{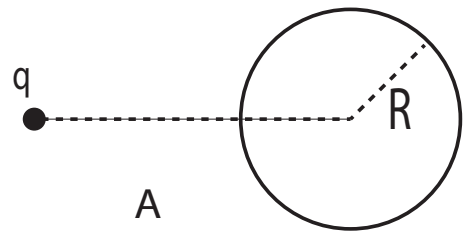}
\caption{This figure represents a spherical wormhole with the radius $R$. A point charge $q$ is placed at the radial distance $r=A$ away from the center of the wormhole.
}
\label{point-charge}
\end{figure}

A realistic traversable wormhole requires the presence of exotic fields that can keep it open. In order to avoid unnecessary complications, we consider a toy model which can be solved analytically and is easy to understand. Consider a flat space in spherical coordinates $(r,\theta, \phi)$. If we place a charge, $q$, at the distance $r=A$ from the center of our coordinate system, this charge will create an electromagnetic potential
\begin{eqnarray}
\label{charge-p}
V_{\rm free}(r)=\frac{q}{4\pi\sqrt{A^2+r^2-2A r\cos\theta}} .
\end{eqnarray}
We now introduce a simple model of a wormhole.
Consider two copies of a flat space connected through a spherical wormhole mouth of the radius $R$. The coordinates in these two flat spaces are $(r_1,\theta_1, \phi_1)$ and $(r_2,\theta_2, \phi_2)$ respectively.
Such a configuration is shown in Fig.~\ref{point-charge}. A charge, $q$, is now placed at the distance $r_1=A$ from the center of a wormhole, in the space $(r_1,\theta_1, \phi_1)$.
Let's call the space where the charge is placed the ``other space", in contrast to the copy with coordinates $(r_2,\theta_2, \phi_2)$ where the observer is located, which we will call ``our space". There is no physical charge in ``our space". The wormhole's radius is $R$. Therefore ``our space" and ``other space" cover only the $r_2>R$ and $r_1>R$ regions respectively.

The presence of the wormhole will inevitably change the flat space potential.  Since near the charge the potential is approximately given by Eq.~\eqref{charge-p}, the potential in the whole "other space" may be written as
\begin{eqnarray}
V_1(r_1)=V_{\rm free}(r_1)+\sum_{l=0}^\infty \frac{T_{l}}{r_1^{l+1}}P_l(\cos\theta_1) ,
\end{eqnarray}
where $P_n(x)$ is an n'th-order Legendre function. The corresponding potential in "our space" is
\begin{eqnarray}
V_2(r_2)=\sum_{l=0}^\infty \frac{B_{l}}{r_2^{l+1}}P_l(\cos\theta_2) .
\end{eqnarray}
Here, $T_l$ and $B_l$ are the coefficients in the expansion.
 We can also expand the free potential in terms of the Legendre functions using
\begin{equation}
\frac{1}{\sqrt{1-2xt+t^2}}=\sum^\infty_{n=0}P_n(x)t^n .
\end{equation}

There are many specific wormhole solutions in literature (see eg. \cite{Dai:2018vrw,Khusnutdinov:2002qb,Kanti:2011jz,Kanti:2011yv,Antoniou:2019awm} and also papers cited in the Introduction). For simplicity, we are not dealing with a specific model, but instead capture some general features that most of the wormholes must posses. For example, we will assume that wormhole throat is very short and therefore the potentials should match at the mouth, $r_1=r_2 =R$. Since there is no charge on the surface of the wormhole, the derivative of the potential in radial direction at the wormhole mouth must be continuous too, i.e.
\begin{eqnarray}
\label{boundary1}
V_1(R)&=&V_2(R)\\
\label{boundary2}
\partial_{r_1} V_1(r_1)|_{r_1=R}&=&-\partial_{r_2} V_2(r_2)|_{r_2=R} .
\end{eqnarray}

Comparing the coefficients of $P_l(\cos\theta)$, we find
\begin{eqnarray}
T_l&=&-\frac{q}{4\pi}\frac{1}{2(l+1)}\frac{R^{2l+1}}{A^{l+1}}\\
B_l&=&\frac{q}{4\pi}\frac{2l+1}{2(l+1)}\frac{R^{2l+1}}{A^{l+1}} .
\end{eqnarray}

Since $B_l \neq 0$, we see that there is an apparent potential in "our space", though there is no physical charge in it. Following the standard procedure, an observer in "our space" can draw a Gaussian surface at $r_2\rightarrow \infty$. He finds that the effective charge in our universe is
\begin{equation}
Q_2=\int -\partial_{r_2} V_2 dA_2\vert_{r_2\rightarrow \infty} =\frac{q}{2}\frac{R}{A} ,
\end{equation}
where $dA_2$ is an element of area in "our space". If an observer in "our space" is not aware of the physical charge in the ``other space", he may conclude that the wormhole has a charge $Q_2$.
The induced effect of the physical charge to ``our space" is stronger if the physical charge is placed closer to the wormhole. When the charge is exactly at the wormhole mouth, i.e. $A=R$, the induced charge $Q_2$ become one half of the original charge $q$.

Similarly, an observer in the "other space" can draw a Gaussian surface at $r_1\rightarrow \infty$, and calculate the value for the charge he observes as
\begin{equation}
Q_1=\int -\partial_{r_1} V_1 dA_1\vert_{r_1\rightarrow \infty} =q-\frac{q}{2}\frac{R}{A}.
\end{equation}
Obviously, the Gaussian flux will be conserved, i.e. $Q_1 +Q_2 =q$, only if we include Gaussian surface on both sides.

\begin{figure}
\includegraphics[width=6cm]{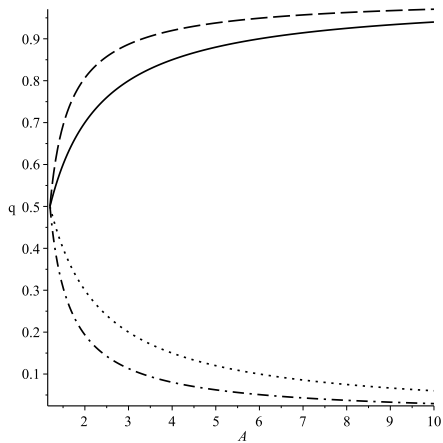}
\caption{ The solid line represents the effective charge, $Q_1$, obtained from the Gaussian surface at infinity in the "other space" (where the original charge was placed). The doted line represents the effective charge, $Q_2$, obtained from the Gaussian surface at infinity in "our space", where the observer is located. The dashed and dotted dashed  lines represent the same quantities $Q_1$ and $Q_2$ but for the Schwarzschild wormhole.  The wormhole radius in that case  is $R= 1.2r_g$.
}
\label{charge-on-bothside}
\end{figure}

 We plot the charges $Q_1$ and $Q_2$ in Fig.~\ref{charge-on-bothside}. When the physical charge is far from the wormhole (i.e. $A\rightarrow \infty$), the effective charge in "our space" is close to $Q_2=0$, while the charge on the other side is $Q_1=q$. As the distance $A$ decreases,
$Q_1$ decreases and $Q_2$ increases. Finally, when the original charged particle is placed exactly at the wormhole mouth, $A=R$, then the effective charges become equal, i.e. $Q_1=Q_2=q/2$. This implies that even before the particle falls into the wormhole, its effect in its world is already diminished. Simultaneously, an observer on the other side of the wormhole (i.e. ``our space") can feel the influence even before the charge crosses over.

\begin{figure}
\includegraphics[width=4cm]{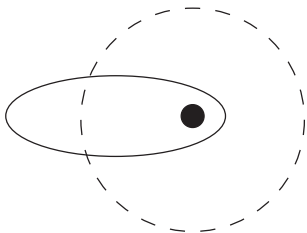}
\caption{The black disk represents a wormhole. A  particle that sources the field can have a circular (dashed line) or an elliptic (solid line) orbit. Other types of orbits are also possible, e.g. hyperbolae or parabolae, but they are not shown in this figure.
}
\label{orbit}
\end{figure}

A physical charge (or particle that sources the field) located in the ``other space" does not have to be at the fixed distance $A$. One can consider a particle orbiting a wormhole in a circular, elliptic or some other type of orbit (see Fig.~\ref{orbit}). For an elliptic orbit, the radius of an orbit, $A$, is not fixed, so the monopole effect will change according to the particle's location. If the orbit is circular, then $A$ is not changing, and one has to consider the multipole effects, primarily a dipole. The effect observed in "our space" will depend on the location of the particle.

\section{ Scalar field and a Schwarzchild wormhole}
\begin{figure}
\includegraphics[width=3cm]{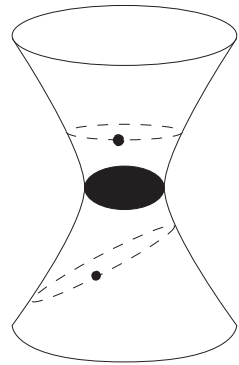}
\caption{A wormhole connects two copies of the Schwarzschild space-time. The source of the Schwarzschild space-time is a black hole with the gravitational radius $r_g$. Stars orbit this wormhole on both sides. If the gravitational field can propagate through the wormhole, then the orbits of the stars will be affected, and will deviate from the standard Schwarzschild orbits.
}
\label{wormhole}
\end{figure}

We now move to a more realistic case, as shown in Fig.~\ref{wormhole}.  We start with the Schwarzschild space-time metric
\begin{equation}
ds^2=-(1-\frac{r_g}{r})dt^2+\frac{1}{1-\frac{r_g}{r}}dr^2+r^2d\Omega
\end{equation}
where $r_g = 2GM$.
Consider now two copies of the Schwarzschild space-time connected through a short throat of radius $R$, which is also the radius of the wormhole mouth. The radius must satisfy $R \ge r_g$, otherwise there would be no distinction between a wormhole and a black hole.
As in the previous section, $r_2$ is the radial coordinate in "our space", while  $r_1$ is the radial coordinate in the ``other space".
Outside of the mouth, i.e. $r_1>R$ and $r_2>R$, the space-time is Schwarzschild on both sides.
These two copies of the Schwarzschild space-time are connected at $r_1=r_2=R$.

We first consider a scalar field propagating in this background.
We place a scalar particle at the distance $r=A$ from the center of a wormhole.
For our purpose, we decompose a single-particle scalar field potential in the basis of the Legendre functions
\begin{eqnarray}
V_{\rm free}({\bf r})=\sum_{l=0}^\infty C_{l} (r)P_l(\cos\theta),
\end{eqnarray}
which satisfies
\begin{equation}
\label{wave}
\Box V_{\rm free}({\bf r})=-\frac{q}{\sqrt{1-r_g/R}}\delta ({\bf r}-{\bf R}).
\end{equation}
Here we use the same definitions as in \cite{Dai:2012ni}. The scalar charge
definition slightly differs from  Eq. 2.2 in \cite{Wiseman:2000rm}. The
difference is the time component of the four-velocity $u^t$, 
which in our case, where the charge is not moving, is just a
constant and can be absorbed in $q$. Since we consider a static case, time derivatives are not included. The wave equation satisfies
\begin{equation}
\frac{1}{r^2}\partial_r \left[(1-r_g/r)r^2\right]\partial_r C_l -\frac{l(l+1)}{r^2}C_l=0 ,
\end{equation}
except at $r=R$ where we have to match the solutions. The $l=0$ solutions is
\begin{eqnarray}
C_0&=& a_0 D_0(r) +b_0 E_0(r)\\
D_0&=&1\\
E_0&=&-\ln (1-\frac{r_g}{r}) ,
\end{eqnarray}
where $a_0$ and $b_0$ are constants.
The $l=1$ solutions is
\begin{eqnarray}
C_1&=& a_1 D_1(r) +b_1 E_1(r)\\
D_1&=&(r-\frac{r_g}{2})\\
E_1&=&-r_g-(r-\frac{r_g}{2})\ln(1-\frac{r_g}{r}) ,
\end{eqnarray}
where $a_1$ and $b_1$ are constants. Since the field is finite and satisfies Eq.~\eqref{wave}, we can write the whole solution as the following.
If $r>A$, then
\begin{eqnarray}
V_{\rm free} = \frac{q}{4\pi r_g}E_0(r)+\frac{12q(A-\frac{r_g}{2})}{4\pi r_g^3} E_1(r)\cos\theta+...
\end{eqnarray}
If $r<A$, then
\begin{eqnarray}
V_{\rm free} = \frac{q}{4\pi r_g}E_0(A)+\frac{12q(A-\frac{r_g}{2})}{4\pi r_g^3}\frac{E_1(A)}{D_1(A)}D_1(r)\cos\theta+...
\end{eqnarray}
Here, $q$ is redefined in accordnace with the asymptotic behavior. As before, we have to find solutions in "our space" and "other space". For this purpose, we decompose the scalar field in the basis of the Legendre functions. The scalar field potential in the "other space" is
 \begin{eqnarray}
V_1 (r_1)&=& V_{\rm free}(r_1)+h_0 E_o(r_1)+h_1E_1(r_1)\cos\theta_1+...
\end{eqnarray}
 The potential in "our space" is
 \begin{eqnarray}
V_2 (r_2)&=& s_0 E_0(r_2)+s_1 E_1(r_2)\cos\theta_2+...
\end{eqnarray}
Exactly at the wormhole mouth, the solution must be continuous and satisfy the conditions in Eqs.~\eqref{boundary1} and \eqref{boundary2}. From this matching, we find the coefficients
\begin{eqnarray}
s_0&=&-h_0=\frac{q}{8\pi r_g}\frac{E_0(A)}{E_0(R)} \\
s_1&=& \frac{6q(A-\frac{r_g}{2})}{4\pi r_g^3}\frac{E_1(A)}{D_1(A)} \Big(\frac{D_1(R)}{E_1(R)}-\frac{D'_1(R)}{E'_1(R)}\Big)\\
h_1&=& -\frac{6q(A-\frac{r_g}{2})}{4\pi r_g^3}\frac{E_1(A)}{D_1(A)} \Big(\frac{D_1(R)}{E_1(R)}+\frac{D'_1(R)}{E'_1(R)}\Big)
\end{eqnarray}

Fig.~\ref{charge-on-bothside} shows the scalar field charges on both sides of the wormhole in the Schwarzschild case. The effect is similar to the electromagnetic case but less pronounced. As the wormhole radius approaches the Schwarzschild horizon, $R \rightarrow r_g$, then $E_0(R)$, $E_1(R)$ and $E'(R)$ approach infinity. In that limit, the scalar field cannot pass the wormhole throat/mouth, $h_0=h_1=s_0=s_1=0$, unless $A=R$. This is not surprising, because in this case the wormhole would not be traversable. This effect is very similar to the black hole no hair theorems.

\section{ Gravity and a Schwarzschild wormhole}
We now move to the gravitational force.
Gravitational perturbations in the Schwarzschild background have been extensively studied \cite{1966PhRv..146..938P,Zerilli:1971wd,Garat:1999vr,Detweiler:2003ci,Barack:2005nr,Chen:2016plo}. We focus on the monopole perturbations, since the higher order modes do not have analytic form in asymptotically flat coordinates. The monopole metric perturbations can be written as
\begin{eqnarray}
\label{perturbation1}
h_{tt}&=&\frac{2\mu}{r}\Theta(r-A)+\frac{2\mu}{A}\Theta(A-r)\\
\label{perturbation2}
h_{rr}&=&\frac{2\mu r }{(r-r_g)^2}\Theta(r-A)
\end{eqnarray}
where $\mu$ is the effective mass of the particle that perturbs the metric, while $\Theta(x)$ is the standard Heaviside function.
This is also an approximative solution for our case since we are working in the thin-shell and short-throat wormhole approximation. We will find the concrete form of our perturbations by matching these forms at the the wormhole mouth.
[On can find a similar form of the monopole metric perturbations for example on the bottom line on the page $18$ in \cite{Chen:2016plo}). However, there the
Zerilli gauge was used which is not appropriate in our case, since we will require that $g_{tt}$ is continuous at the shell of the radius $R$ (the radius of the wormhole mouth). Continuous $g_{tt}$ and $g_{rr}$ can be found in the Lorentz gauge (see for example Fig.~1 in \cite{Barack:2005nr}). Therefore our expression differs from that in \cite{Chen:2016plo} in the second term on the right hand side of Eq.~(\ref{perturbation1}).]

 As we mentioned, we require that $h_{tt}$ is continuous at $r=R$, so the time variable is the same inside and outside of the shell $r=R$. The metric perturbations $h_{tt}$ and $h_{rr}$ are not completely independent, since they both depend on the mass of the particle.  All the other metric perturbations are zero in this case.

We can now write the perturbations in the "other space" as
\begin{eqnarray}
h^{oth}_{tt}(r_1)&=&h_{tt}(r_1)+\frac{2a_{tt}}{r_1}\\
h^{oth}_{rr}(r_1)&=&h_{rr}(r_1)+\frac{2a_{rr} r_1 }{(r_1-r_g)^2} .
\end{eqnarray}
From eq. \eqref{perturbation1} and \eqref{perturbation2}, we see that $a_{tt}$ and $a_{rr}$ are not completely independent. They are both equal to an effective mass in the ``other space", $a_{tt}=a_{rr}=\mu^{oth}$. The perturbations in "our space" are
\begin{eqnarray}
h^{our}_{tt}(r_2)&=&\frac{2b_{tt}}{r_2}\\
h^{our}_{rr}(r_2)&=&\frac{2b_{rr} r_2 }{(r_2-r_g)^2} .
\end{eqnarray}
Again $b_{tt}$ and $b_{rr}$ are not completely independent, and they are both equal to an effective mass in our space, $b_{tt}=b_{rr}=\mu^{our}$. We can find $a_{tt}$, $a_{rr}$, $b_{tt}$ and $b_{rr}$ from the continuity condition ($h^{our}_{tt}(R)=h^{oth}_{tt}(R)$ and $\partial_{r_2}h^{our}_{tt}\vert_{r_2=R}=\partial_{r_1}h^{our}_{tt}\vert_{r_1=R}$) as

\begin{eqnarray}
b_{tt}&=&-a_{tt}=\mu \frac{R}{A}\\
b_{rr}&=&-a_{rr}=\mu \frac{R }{A} .
\end{eqnarray}

Since $b_{tt}$ is nonzero, a static observer can feel an additional (or anomalous) acceleration coming from the wormhole. If the observer is far away from the wormhole, this additional acceleration is
\begin{equation}
\label{acceleration}
a\approx-\mu \frac{R}{A}\frac{1}{r_2^2} .
\end{equation}
In contrast with the scalar field case, the effect still exists even if $R=r_g$. 
In the scalar field case, we saw that the scalar field shuts off as $R\rightarrow r_g$ (wormhole mouth approaches the gravitational radius of a black hole).   If it did not shut off, it would directly violate the scalar no-hair theorems for black holes. There are no gravitational no-hair theorems for black holes, so it might be ok that gravitational perturbations do not shut-off as $R\rightarrow r_g$. No information travels from within the horizon out, but the source of perturbations perturbs the horizon, and a test object propagating in vicinity of the horizon on the other side feels these perturbations. However, within the approximations in our paper, it is not quite clear what happens. One would have to do a full time-dependent analysis for a sequence of wormhole geometries as $R\rightarrow r_g$. It is not excluded that one would find some superssion factors as $R\rightarrow r_g$, but at least in our approximation they do not show up.

\section{ Observational signature of a wormhole}
It is widely accepted that Sgr A* harbors a super massive black hole at the center of our galaxy \cite{1977ApJ...218L.103W,1996ApJ...472..153G,2007gsbh.book.....M,Ghez:2008ms,Schodel:2002vg,2009ApJ...707L.114G,Gillessen:2008qv,2007ApJ...659..378R}. Sgr A* contains mass of $M_\bullet=4\times 10^6 M_\odot$, with the corresponding Schwarzschild radius of $r_g=0.084$AU. While many potential effects may affect the orbits the stars that orbit this black hole \cite{Iorio:2010hw}, we would like to explore perhaps the most interesting possibility that Sgr A* might be a wormhole. In this case, stars orbiting around it in the ``other space" should affect stars' orbits in our universe.

We choose the star $S2$, which orbits the radio source Sgr A*. Its mass is about $14$ solar masses, with an orbital period of $15.9$ years, a semi-major axis of $1031.69$ AU, and orbit ellipticity of $e=0.8831$  \cite{2009ApJ...707L.114G}.
Since we  calculated only the monopole contribution, we consider an object orbiting on the other side of the wormhole with the periapsis radius $r_p$ and apoapsis radius $r_a$. From Eq.~\eqref{acceleration}, the contribution from the monopole perturbation causes the acceleration variation
\begin{equation} \label{deltaa}
\Delta a =\mu \left(\frac{R}{r_p}-\frac{R}{r_a}\right)\frac{1}{r_2^2} .
\end{equation}
If the orbit of an object on the other side of  the wormhole's  is elongated so that $r_a\gg r_p$, then we can approximate
\begin{equation}
\Delta a =\mu \frac{R}{r_p}\frac{1}{r_2^2} .
\end{equation}
To obtain definite values, we set the wormhole mouth to be of the order of the black hole horizon, i.e. $R\approx r_g$, and the orbit for the $S2$ star, $r_2\approx 1000$AU, to find the constraints on the detectable values of $\mu$ and $r_p$.

The total acceleration of the star $S2$ in orbit is $1.5$m/s$^2$, and mainly comes from the supermassive black hole. 
This acceleration has been measured with the precision of $4\times 10^{-4} m/s^2$, with two years (1997-1999) of data\cite{Iorio:2010hw,Ghez:2000ay}. With $20$ years data, it should be possible to achieve the precision of $2\times 10^{-5} m/s^2$ \cite{Iorio:2010hw,2009ApJ...700..137R}. This can be further improved to $10^{-6} m/s^2$ if the velocity uncertainty is reduced to $2 km/s$\cite{Iorio:2010hw,Ghez:2008ms,Gould:2002hf}. The best available data today (see e.g. Table 8 in \cite{20years}) is still shy (though not unreasonably far) from this precision.

Note that what we calculate in our Eq.~(\ref{deltaa}) is {\it acceleration variation} of the star S2 due to an elliptic orbit of the star on the other side perturbing the metric.
These variations come on top of the constant acceleration that comes from the central black hole.  
With good enough precision, we should be able to detect or exclude this variable anomalous acceleration. Of course, these variations can possibly be produced by some other sources, for example by other smaller black holes in vicinity of S2. Then, more careful modeling would be required to distinguish between different options.  
 
\begin{figure}
\includegraphics[width=6cm]{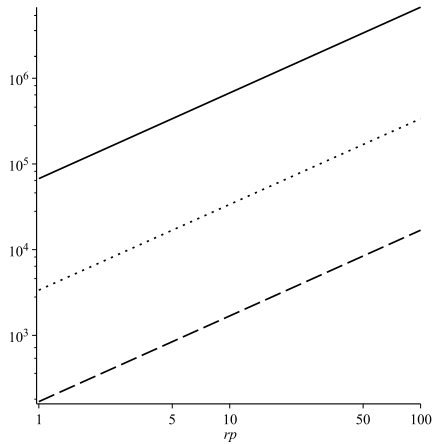}
\caption{We plot the constraints on the mass $\mu$ and the periapsis radius $r_p$ of a hypothetical star that orbits Sgr A* on the ``other side" and perturbs the orbit of the S2 star on our side.
The black, doted, and dashed lines represent the constraints with acceleration precision of the star S2 of $4\times 10^{-4} m/s^2$, $2\times 10^{-5} m/s^2$, and $10^{-6} m/s^2$ respectively. The regions above the  lines rule out a wormhole explanation. The x-axis has units of $r_g$. The y-axis has units of $M_\odot$. The bottom line probes the most reasonable parameter space - a few solar masses star orbiting around Sgr A* at the distance of a few gravitational radii.
}
\label{constraint}
\end{figure}

Fig~\ref{constraint} shows the regions which are ruled out with the recent acceleration precision and potential acceleration precision that can be obtained in future. If Sgr A* is a wormhole with heavy stars orbiting it on the ``other side", we can definitely see the effects in the near future. In particular, with acceleration precision of $10^{-6} m/s^2$, a few solar masses star orbiting around Sgr A* at the distance of a few gravitational radii would leave detectable imprint on the orbit of the S2 star on our side of the wormhole. 

 A different test on the hot gas motion near Sgr A* was proposed in  \cite{Li:2014coa}. This is important since different tests may be combined to check whether some black hole candidate is actually a wormhole.

In addition, one can also look for the same effect in the binary systems of a black hole and a star.  A deviation of motion of a star could be a hint of the existence of a wormhole, if it is consistent with perturbations we derived here. Finally, if the monopole contribution is insufficient, one can consider dipole or higher multipole effects in order to extract more stringent constrains or definite predictions.

\begin{acknowledgments}
We are very thankful to K. Thorne and J. Mas for very useful discussions.
D.C Dai was supported by the National Science Foundation of China (Grant No. 11433001 and 11775140), National Basic Research Program of China (973 Program 2015CB857001) and  the Program of Shanghai Academic/Technology Research Leader under Grant No. 16XD1401600. DS was partially supported by the US National Science Foundation, under Grant No. PHY-1820738.

\end{acknowledgments}


\begin{thebibliography}{99}

\bibitem{ER} A. Einstein and N. Rosen, Phys. Rev. {\bf 48}, 73 (1935).

\bibitem{geons} J. A. Wheeler, 
Phys.\ Rev.\ {\bf 97}, 511 (1955).

\bibitem{wheeler} J. A. Wheeler, {\it Geometrodynamics}, Academic, New York, 1962.

\bibitem{Baum:1984mc}
  E.~Baum,
  Phys.\ Lett.\  {\bf 133B}, 185 (1983).

\bibitem{Hawking:1984hk}
  S.~W.~Hawking,
  Phys.\ Lett.\  {\bf 134B}, 403 (1984).

\bibitem{Coleman:1988tj}
  S.~R.~Coleman,
  Nucl.\ Phys.\ B {\bf 310}, 643 (1988).


\bibitem{gibbons}
G. W. Gibbons and S. W. Hawking (editors), {\it Euclidean Quantum Gravity}, World Scientific, 1993.

\bibitem{Morris:1988cz}
  M.~S.~Morris and K.~S.~Thorne,
  Am.\ J.\ Phys.\  {\bf 56}, 395 (1988).

\bibitem{Morris:1988tu}
  M.~S.~Morris, K.~S.~Thorne and U.~Yurtsever,
  Phys.\ Rev.\ Lett.\  {\bf 61}, 1446 (1988).


\bibitem{visser} M. Visser, {\it Lorentzian Wormholes: From Einstein to Hawking}, AIP Press, New York, 1995


\bibitem{Maldacena:2013xja}
  J.~Maldacena and L.~Susskind,
  Fortsch.\ Phys.\  {\bf 61}, 781 (2013)
  [arXiv:1306.0533 [hep-th]].


\bibitem{Gratus:2019ned}
  J.~Gratus, P.~Kinsler and M.~W.~McCall,
  Found. Phys. 49, (2019) [published online April 4]
  doi:10.1007/s10701-019-00251-5
  [arXiv:1904.04103 [gr-qc]].


\bibitem{Bronnikov:2019gsr}
  K.~A.~Bronnikov, S.~V.~Bolokhov and M.~V.~Skvortsova,
  arXiv:1903.09862 [gr-qc].

\bibitem{Bronnikov:2018vbs}
  K.~A.~Bronnikov,
  Particles {\bf 1}, no. 1, 56 (2018)
  doi:10.3390/particles1010005
  [arXiv:1802.00098 [gr-qc]].

\bibitem{Ovgun:2018xys}
  A.~?vgün, K.~Jusufi and İ.~Sakallı,
  Phys.\ Rev.\ D {\bf 99}, no. 2, 024042 (2019)
  doi:10.1103/PhysRevD.99.024042
  [arXiv:1804.09911 [gr-qc]].

\bibitem{Jusufi:2017mav}
  K.~Jusufi and A.~?vgün,
  Phys.\ Rev.\ D {\bf 97}, no. 2, 024042 (2018)
  doi:10.1103/PhysRevD.97.024042
  [arXiv:1708.06725 [gr-qc]].

\bibitem{Jusufi:2016leh}
  K.~Jusufi,
  Eur.\ Phys.\ J.\ C {\bf 76}, no. 11, 608 (2016)
  doi:10.1140/epjc/s10052-016-4456-3
  [arXiv:1607.04070 [gr-qc]].


\bibitem{Krasnikov:2008kr}
  S.~Krasnikov,
  Class.\ Quant.\ Grav.\  {\bf 25}, 245018 (2008)
  doi:10.1088/0264-9381/25/24/245018
  [arXiv:0802.1358 [gr-qc]].

\bibitem{Dai:2018vrw}
  D.~C.~Dai, D.~Minic and D.~Stojkovic,
  Phys.\ Rev.\ D {\bf 98}, no. 12, 124026 (2018)
  doi:10.1103/PhysRevD.98.124026
  [arXiv:1810.03432 [hep-th]].

\bibitem{Khusnutdinov:2002qb}
  N.~R.~Khusnutdinov and S.~V.~Sushkov,
  Phys.\ Rev.\ D {\bf 65}, 084028 (2002)
  doi:10.1103/PhysRevD.65.084028
  [hep-th/0202068].

\bibitem{Kanti:2011jz}
  P.~Kanti, B.~Kleihaus and J.~Kunz,
  Phys.\ Rev.\ Lett.\  {\bf 107}, 271101 (2011)
  doi:10.1103/PhysRevLett.107.271101
  [arXiv:1108.3003 [gr-qc]].

\bibitem{Kanti:2011yv}
  P.~Kanti, B.~Kleihaus and J.~Kunz,
  Phys.\ Rev.\ D {\bf 85}, 044007 (2012)
  doi:10.1103/PhysRevD.85.044007
  [arXiv:1111.4049 [hep-th]].

\bibitem{Antoniou:2019awm}
  G.~Antoniou, A.~Bakopoulos, P.~Kanti, B.~Kleihaus and J.~Kunz,
  arXiv:1904.13091 [hep-th].

\bibitem{Dai:2012ni} 
  D.~C.~Dai and D.~Stojkovic,
  Phys.\ Rev.\ D {\bf 86}, 084034 (2012)
  doi:10.1103/PhysRevD.86.084034
  [arXiv:1209.3779 [hep-th]].
  
\bibitem{Wiseman:2000rm} 
  A.~G.~Wiseman,
  Phys.\ Rev.\ D {\bf 61}, 084014 (2000)
  doi:10.1103/PhysRevD.61.084014
  [gr-qc/0001025].

  \bibitem[Peters(1966)]{1966PhRv..146..938P} Peters, P.~C.\ 1966, Physical Review, 146, 938

\bibitem{Zerilli:1971wd}
  F.~J.~Zerilli,
  Phys.\ Rev.\ D {\bf 2}, 2141 (1970).
  doi:10.1103/PhysRevD.2.2141

\bibitem{Garat:1999vr}
  A.~Garat and R.~H.~Price,
  Phys.\ Rev.\ D {\bf 61}, 044006 (2000)
  doi:10.1103/PhysRevD.61.044006
  [gr-qc/9909005].

\bibitem{Detweiler:2003ci}
  S.~L.~Detweiler and E.~Poisson,
  Phys.\ Rev.\ D {\bf 69}, 084019 (2004)
  doi:10.1103/PhysRevD.69.084019
  [gr-qc/0312010].


\bibitem{Barack:2005nr}
  L.~Barack and C.~O.~Lousto,
  Phys.\ Rev.\ D {\bf 72}, 104026 (2005)
  doi:10.1103/PhysRevD.72.104026
  [gr-qc/0510019].

\bibitem{Chen:2016plo}
  J.~E.~Thompson, B.~F.~Whiting and H.~Chen,
  Class.\ Quant.\ Grav.\  {\bf 34}, no. 17, 174001 (2017)
  doi:10.1088/1361-6382/aa7f5b
  [arXiv:1611.06214 [gr-qc]].

\bibitem{Ghez:2008ms}
  A.~M.~Ghez {\it et al.},
  Astrophys.\ J.\  {\bf 689}, 1044 (2008)
  doi:10.1086/592738
  [arXiv:0808.2870 [astro-ph]].


\bibitem[Gillessen et al.(2009)]{2009ApJ...707L.114G} Gillessen, S., Eisenhauer, F., Fritz, T.~K., et al. 2009, apjl, 707, L114



\bibitem{Gillessen:2008qv}
  S.~Gillessen, F.~Eisenhauer, S.~Trippe, T.~Alexander, R.~Genzel, F.~Martins and T.~Ott,
  Astrophys.\ J.\  {\bf 692}, 1075 (2009)
  doi:10.1088/0004-637X/692/2/1075
  [arXiv:0810.4674 [astro-ph]].


\bibitem[Reid et al.(2007)]{2007ApJ...659..378R} Reid, M.~J., Menten, K.~M., Trippe, S., et al.\ 2007, \apj, 659, 378.



\bibitem[Wollman et al.(1977)]{1977ApJ...218L.103W} Wollman, E.~R., Geballe, T.~R., Lacy, J.~H., Townes, C.~H., \& Rank, D.~M.\ 1977, apjl, 218, L103


\bibitem[Genzel et al.(1996)]{1996ApJ...472..153G} Genzel, R., Thatte, N., Krabbe, A., Kroker, H., \& Tacconi-Garman, L.~E.\ 1996, apj, 472, 153

\bibitem{Schodel:2002vg}
  R.~Schodel {\it et al.},
  Nature
  [Nature {\bf 419}, 694 (2002)]
  doi:10.1038/nature01121
  [astro-ph/0210426].

  \bibitem[Melia(2007)]{2007gsbh.book.....M} Melia, F.\ 2007, The Galactic Supermassive Black Hole by Fulvio Melia.~Princeton University Press, 2007.,



\bibitem{Iorio:2010hw}
  L.~Iorio,
  Mon.\ Not.\ Roy.\ Astron.\ Soc.\  {\bf 411}, 453 (2011)
  doi:10.1111/j.1365-2966.2010.17701.x
  [arXiv:1008.1720 [gr-qc]].




\bibitem{Ghez:2000ay}
  A.~Ghez, M.~Morris, E.~E.~Becklin, T.~Kremenek and A.~Tanner,
  Nature {\bf 407}, 349 (2000)
  doi:10.1038/35030032, 10.1038/407349a
  [astro-ph/0009339].



\bibitem[Reid et al.(2009)]{2009ApJ...700..137R} Reid, M.~J., Menten, K.~M., Zheng, X.~W., et al.\ 2009, apj, 700, 137





\bibitem{Gould:2002hf}
  A.~Gould,
  Astrophys.\ J.\  {\bf 583}, 765 (2003)
  Erratum: [Astrophys.\ J.\  {\bf 607}, 653 (2004)]
  doi:10.1086/345446, 10.1086/383344
  [astro-ph/0208004].






 \bibitem{20years}
 Boehle, A.; Ghez, A. M.; Schödel, R.; Meyer, L.; Yelda, S.; Albers, S.; Martinez, G. D.; Becklin, E. E.; Do, T.; Lu, J. R.; Matthews, K.; Morris, M. R.; Sitarski, B.; Witzel, G.
  The Astrophysical Journal, Volume 830, Issue 1, article id. 17, 23 pp. (2016).
  [arXiv:1607.05726[astro-ph.GA]]


  \bibitem{Li:2014coa}
  Z.~Li and C.~Bambi,
  Phys.\ Rev.\ D {\bf 90}, 024071 (2014)
  doi:10.1103/PhysRevD.90.024071
  [arXiv:1405.1883 [gr-qc]].




\end{thebibliography}
\end{document}